\documentclass[aps,pra,groupedaddress,twocolumn,showpacs]{revtex4-1}

\usepackage{amssymb}
\usepackage{graphicx}
\usepackage{amsmath}

\newcommand{\Nt}{ {N_\textrm{t}} }
\newcommand{\qn}{ ^\textrm{n} }
\newcommand{\dx}{ \dot{x} }
\newcommand{\ket}[1]{\ensuremath{|#1\rangle}}

\newcommand{\pref}[1]{(\ref{#1})}

\newcommand{\Fref}[1]{Figure~\ref{#1}}
\newcommand{\fref}[1]{Fig.~\ref{#1}}

\newcommand{\Eref}[1]{Equation~\pref{#1}}
\newcommand{\eref}[1]{Eq.~\pref{#1}}

\begin{document}
\title{Solving the transport without transit quantum paradox \\ of the spatial adiabatic passage technique}

\author{A.~Benseny,$^{1}$ J.~Bagud\`a,$^{1}$ X.~Oriols,$^{2}$ and J.~Mompart$^{1}$}

\affiliation{$^1$Departament de F\'{i}sica, 
Universitat Aut\`{o}noma de Barcelona, E-08193 Bellaterra, Spain } 

\affiliation{$^2$Departament d'Enginyeria Electr\`onica, 
Universitat Aut\`{o}noma de Barcelona, E-08193 Bellaterra, Spain } 

\begin{abstract}
We discuss and solve the transport without transit quantum paradox recently introduced in the context of the adiabatic transport of a single particle or a Bose--Einstein condensate between the two extreme traps of a triple-well potential.
To this aim, we address the corresponding quantum dynamics in terms of Bohmian trajectories and show that transport always implies transit through the middle well, in full agreement with the quantum continuity equation.
This adiabatic quantum transport presents a very counterintuitive effect: by slowing down the total time duration of the transport process, ultra-high Bohmian velocities are achieved such that, in the limit of perfect adiabaticity, relativistic corrections are needed to properly address the transfer process while avoiding superluminal matter wave propagation.
\end{abstract}

\date{\today }

\maketitle

Classical laws govern all of our daily physical situations providing us with an intuition on what is physically feasible.
It is not surprising then that a wide collection of quantum and relativistic phenomena escape from our `limited' classical intuition.
Very often, the explanation of some puzzling non-classical effects has led to apparent contradictions that defy intuition such as, to cite only a few, the well known Schr\"odinger cat and  Einstein--Podolsky--Rosen quantum paradoxes, or the tunnel and twin paradoxes in the relativistic realm.
The detailed analysis of such paradoxes has allowed physicists to achieve a deeper understanding on the background of both quantum mechanics and relativity which, in turn, has strongly influenced modern research in quantum information and cosmology, among other fields.

A novel quantum paradox named \textit{transport without transit} (TWT)~\cite{TWT} was recently formulated in the context of the transport of a Bose--Einstein condensate (BEC) between the two extreme traps of a triple-well potential by means of the matter wave analog of the quantum optical stimulated Raman adiabatic passage (STIRAP) technique~\cite{STIRAP}.
This spatial adiabatic transport technique has been studied for both single particles~\cite{TLAO,Opatrny-Das,MWS-Atom,Hole} and BECs~\cite{MWS-BEC,TWT} and consists in adiabatically following an energy eigenstate of the system, the so-called spatial dark state, that only populates the vibrational ground states of the two outer wells and presents at all times a node in the central region.
Thus, the matter wave is transferred between the outer traps, but maintaining an arbitrarily close to zero population in the middle region during the whole evolution~\cite{TWT,Opatrny-Das}.
From this observation,  M.~Rab {\it et al.} have concluded that quantum-mechanically it is possible to transport matter \textit{directly} from the left trap to the right one, without transiting the center region, and formulated the TWT paradox~\cite{TWT}:
``Classically it is impossible to have transport without transit, i.e., if the points 1, 2, and 3 lie sequentially along a path then an object moving from 1 to 3 must, at some time, be located at 2. For a quantum particle in a three-well system it is possible to transport the particle between wells 1 and 3 such that the probability of finding it at \textit{any time} in the classical accessible state in well 2 is negligible.''
Clearly, quantum TWT is in contradiction with the continuity equation that derives from both the single-particle Schr\"odinger equation and the Gross-Pitaevskii equation (GPE), that governs the ultracold BEC dynamics in the mean field approximation~\cite{GrossPitaevskii}.

Bohmian mechanics~\cite{Bohm}, while being equivalent to standard quantum mechanics when averaged over the complete set of initial conditions, provides a very clear physical picture of the quantum continuity equation.
It is worth noting that average trajectories on a double slit experiment constructed from weak measurements of photon momenta have been experimentally observed recently, and are identical to the corresponding Bohmian trajectories~\cite{Steinberg}.
In this work we will numerically investigate the adiabatic transport of a BEC in a triple-well potential in terms of Bohmian trajectories to demonstrate that matter wave transport always implies transit.
We will show that by slowing down the transport process, Bohmian velocities in the middle well increase with no apparent limit.
The appearance of superluminal velocities is ``an artifact of using the nonrelativistic Schr\"odinger equation''~\cite{Superluminal}, and thus in the limit of perfect adiabaticity relativistic corrections would be needed to properly describe the system dynamics.
Ultimately, the origin of the TWT paradox is the incorrect use of the (nonrelativistic) Schr\"odinger equation in the adiabatic limit.

The GPE governing the dynamics of a BEC reads:
\begin{equation} \label{eq.GPE}
i \hbar  \frac{\partial \psi (\vec{r},t)}{\partial t} = \left[ - \frac{\hbar^2}{2m}\nabla^2+V(\vec{r},t)+g \left| \psi (\vec{r},t) \right|^2 \right] \psi (\vec{r},t) ,
\end{equation}
where $\psi$ is the wavefunction of the BEC of mass $m$, $V$ is the trapping potential and $g$ governs the non-linear interaction between the atoms~\cite{GrossPitaevskii}.
For $g=0$ we recover the standard Schr\"odinger equation that deals with the dynamics of a single atom or a non-interacting BEC.
In Bohmian mechanics~\cite{Bohm}, $\psi(\vec{r},t)$ guides an ensemble of $\Nt \to \infty$ trajectories, $\{ \vec{r}_k[t] \}$, whose velocity is given by
\begin{align} \label{eq.velocity}
\dot{\vec{r}}_k[t] &= \left.v(\vec{r},t)\right|_{\vec{r}=\vec{r}_k[t]},  \\
\vec{v}(\vec{r},t) &= \frac{\vec{J}(\vec{r},t)}{|\psi(\vec{r},t)|^2} = \frac{1}{m} \vec{\nabla} S(\vec{r},t).
\end{align}
$S$ is the phase of the wavefunction and $\vec{J}$ is the probability current density.
The trajectory ensemble reproduces the probability and current densities,
\begin{align}
|\psi(\vec{r},t)|^2 &= \frac{1}{\Nt} \sum^{\Nt}_{k=1} \delta(\vec{r}-\vec{r}_k[t]), \label{eq.R2traj} \\
\vec{J}(\vec{r}, t) &= \frac{1}{\Nt} \sum^{\Nt}_{k=1} \dot{\vec{r}}_k[t] \delta(\vec{r}-\vec{r}_k[t]), \label{eq.Jtraj}
\end{align}
and thus, averages over the trajectory ensemble are equivalent to mean values obtained from the wavefunction.
Furthermore, by inserting the wavefunction in polar form, i.e., $\psi=|\psi| e^{iS/\hbar}$, in the GPE and separating real and imaginary parts we are led to the equations:
\begin{align}
\label{eq.QHJ}
\frac{\partial S}{\partial t} &= - \left( V + g |\psi|^2 + \frac{1}{2} m \vec{v}\,^2 - \frac{\hbar^2}{2 m}\frac{\nabla^2 |\psi|}{|\psi|} \right), \\
\label{eq.continuity}
\frac{\partial |\psi|^2}{\partial t} &= - \vec{\nabla} \cdot  \left( |\psi|^2 \vec{v} \right) .
\end{align}
\Eref{eq.QHJ} is the so-called quantum Hamilton--Jacobi equation, analogous to its classical counterpart, but with an additional term, the quantum potential, that accounts for the quantum features of the system.
\Eref{eq.continuity} is the quantum continuity equation for the condensate density that ensures that transport is always carried out in a continuous fashion and that Eqs.~(\ref{eq.R2traj}--\ref{eq.Jtraj}) are verified at all times~\cite{Bohm}.

It has been shown that the TWT paradox appears in both the 1D and the full 3D cases~\cite{TWT}.
Thus, to ease the computational burden we will work with the straightforward reduction of Eqs.~(\ref{eq.GPE}--\ref{eq.continuity}) to one spatial dimension.
Following Ref.~\cite{TWT}, we consider a triple-well potential consisting of a harmonic trap of frequency $\omega_x$ (and ground state width $\alpha^{-1} = \sqrt{\hbar/m \omega_z}$) where two Gaussian barriers of heights $V_{12}$ and $V_{23}$ have been added at positions $\mp x_0$, giving a total potential of the form:
\begin{equation} \label{eq.potential}
V(x,t) = \frac{m}{2} \omega_x^2 x^2 + V_{12}(t)e^{-\frac{(x+x_0)^2}{2\sigma^2}}+ V_{23}(t)e^{-\frac{(x-x_0)^2}{2\sigma^2}}
\end{equation}
The barriers, created for instance with two blue-detuned lasers, will be considered initially large enough to inhibit tunneling such that three distinct traps are well defined.
The eventual lowering of the barriers would allow the BEC, located initially in the ground state of the left trap, to tunnel to the other trapping regions. 

The adiabatic transport of the BEC from the left to the right trap is achieved by following an energy eigenstate of the system, the dark state, that only involves the ground states of the left and right traps, namely $\ket{L}$ and $\ket{R}$, and reads $\ket{D(\theta)}=\cos \theta \ket{L} - \sin \theta \ket{R}$, where the mixing angle $\theta$ is defined as $\tan \theta = \Omega_1 / \Omega_2 $, being $\Omega_1$ ($\Omega_2$) the tunneling rate between left and middle (middle and right) traps~\cite{TLAO}. 
By lowering the barriers in a counter-intuitive fashion, i.e., favoring first the tunneling between the middle and right traps and then the tunneling between the left and middle traps, $\ket{D(\theta)}$ transforms from $\ket{L}$ to $\ket{R}$~\cite{TLAO}.
With this in mind, the temporal profile of the potential barriers height is taken to be:
\begin{align}
\label{eq.V23}
V_{23}(t) &=
\begin{cases}
V_{\rm max} & t \leq 0 \\
(V_{\rm max} - V_{\rm min}) \left(\dfrac{2t}{t_{\rm p}} - 1\right)^4 + V_{\rm min}  & 0 < t < t_{\rm p} \\
V_{\rm max} & t \geq t_{\rm p}
\end{cases} , \\
\label{eq.V12}
V_{12}(t) &= V_{23}(t-t_{\rm d})  ,
\end{align}
being $t_{\rm d}$ the time delay between the pulsing of $V_{23} (t)$ and $V_{12} (t)$, giving a total adiabatic transport time $T = t_{\rm p} + t_{\rm d}$.
For the numerical simulations, we take parameter values similar to Ref.~\cite{TWT}:
$V_{\rm min} = 5 \hbar \omega_x$,
$V_{\rm max} = 10^3 \hbar \omega_x$,
$\sigma = 0.16 \alpha^{-1}$ and
$x_0 = 0.48 \alpha^{-1}$.
We will simulate the process for different values of $g$ to study the effect of the nonlinear interaction.
We will also analyze the effects of slowing down the process, by varying the pulsing time $t_{\rm p}$, even though for $\omega_x t_{\rm p} < 1000$ the dynamics is not adiabatic enough and the transport breaks down.
The delay $t_{\rm d}$ is in all cases taken to be $0.15 t_{\rm p}$.

\begin{figure}
\centerline{\includegraphics[width=\columnwidth]{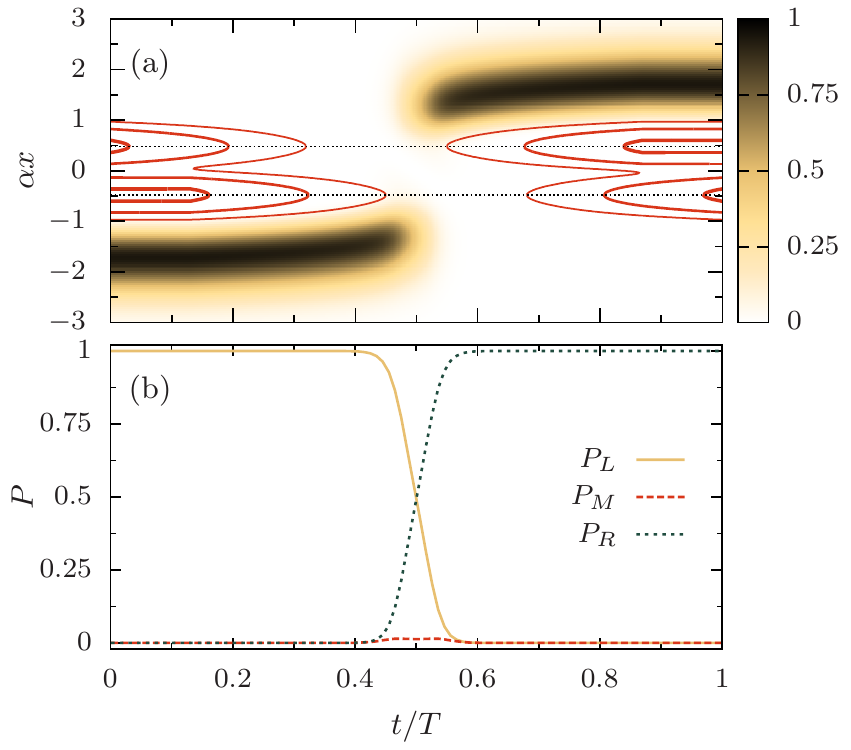}}
\caption{
\label{fig.wfxt}
(color online)
Adiabatic BEC transport for $g=0$ and $\omega_x t_{\rm p} = 5000$.
(a) Atomic probability density $|\psi(x,t)|^2$, 
barrier positions $x=\pm x_0$ (dotted lines) and 
potential contour lines for $V(x,t) = 10$ $\hbar \omega_x$, 100 $\hbar \omega_x$ and 750 $\hbar \omega_x$ (from thinner to thicker).
(b) Integrated populations $P_i=\int_{{\mathcal C}_i} |\psi(x,t)|^2 dx$ with $i=L,\, M,\, R$ in the regions ${\mathcal C}_L = (-\infty, \,-x_0)$, ${\mathcal C}_M = (-x_0, \,x_0)$ and ${\mathcal C}_R = (x_0, \, \infty)$.
}
\end{figure}

We show in \fref{fig.wfxt}(a) the probability distribution during the BEC transport with the barrier heights from Eqs.~(\ref{eq.V23}--\ref{eq.V12}).
Indeed, when the potential barriers are lowered it appears as if the condensate is vanishing from the left region and appearing in the right region without transiting the space between the barriers.
This is validated by \fref{fig.wfxt}(b), that shows that while the BEC population is completely transfered from the left to the right region, the middle region remains unpopulated the entire time, due to the fact that the dark state that the system is (ideally) following presents a node in this region.

\begin{figure}
\centerline{\includegraphics[width=\columnwidth]{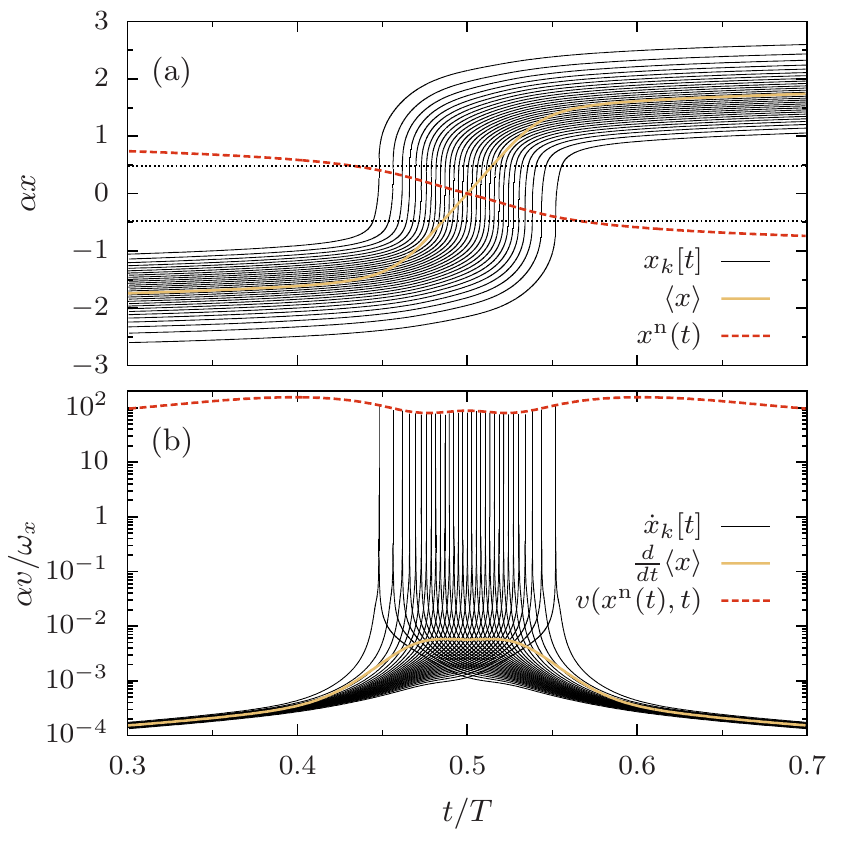}}
\caption{
\label{fig.traj}
(color online)
(a) Bohmian trajectories $x_k[t]$ associated to the evolution shown in \fref{fig.wfxt}(a) (black lines),
atomic mean position (solid line),
dark state node position $x\qn(t)$ (dashed line),
and barrier positions $x=\pm x_0$ (dotted lines).
(b) Bohmian velocities $\dx_k[t]$ of the trajectories in (a) (black lines),
velocity of the atomic mean position (solid line),
and Bohmian velocity at the node position (dashed line).
}
\end{figure}

To elucidate the transport process we have computed the Bohmian trajectories associated to this evolution, that are shown in \fref{fig.traj}(a).
The trajectories, initially distributed according to $|\psi(x,t=0)|^2$, follow the wavefunction at all times, starting in the left trap and ending in the right one.
All the trajectories cross the middle region, demonstrating the non-existence of transport without transit, but in order to keep a low population of the middle region by minimizing the time spent there, they are forced to increase their velocity~\cite{Hole}.
We compute the dark state node position, $x\qn(t)$, as the position in the central region where the population is minimum, and see that the trajectories accelerate when they get closer to it, reaching velocities much larger than the mean wavepacket velocity, see \fref{fig.traj}(b).
Note that each trajectory velocity peaks at a different time, $t\qn_k$, depending on its initial position inside the wavefunction, but all trajectories achieve similar maximum velocities.
$t\qn_k$ corresponds to the time at which the $k$-th trajectory crosses the dark state node, i.e., $x_k[t\qn_k]=x\qn(t\qn_k)$, since the trajectories peak velocities correspond to the Bohmian velocity at the node, see dashed line in \fref{fig.traj}(b).

Since the transport is performed in a finite time, the dynamic state $\psi(x,t)$ will not be exactly equal to the dark state and will present some small population from other states giving a non-zero atomic population at the node position during the evolution.
If the evolution is made more adiabatic by performing the process more slowly, the dynamic state will remain closer to the dark state and the node population will be smaller.
This will result in an increase of the trajectories velocity at the node (cf. \eref{eq.velocity}),
presenting a surprising effect: by slowing down the total matter wave STIRAP sequence it is possible to achieve sudden trajectory accelerations yielding ultra-high velocities.

In order to obtain a quantitative argument, we compute the ensemble average of the velocity that the trajectories achieve at the node, $\langle v^\textrm{max} \rangle$.
By using Eqs.~(\ref{eq.velocity}--\ref{eq.Jtraj}) and taking into account that $\dx_k[t] \geq 0$ and $\dx\qn(t) \leq 0$ (cf. \fref{fig.traj}(a)) such average can be rewritten as:
\begin{align}
\label{eq.vmax}
\langle v^\textrm{max} \rangle
&= \frac{1}{\Nt} \sum_{k=1}^\Nt \dx_k[t\qn_k] \nonumber \\
&= \frac{1}{\Nt} \sum_{k=1}^\Nt \int_0^T \delta(t-t\qn_k) v(x_k[t],t) dt \nonumber \\
&= \frac{1}{\Nt} \sum_{k=1}^\Nt \int_0^T \delta(x_k[t] - x\qn(t)) \nonumber \\
& \qquad \qquad \qquad \times v(x_k[t],t) [v(x_k[t],t) -  \dx\qn(t)] dt \nonumber \\
&= \int_0^T  \! \left [ \frac{J(x\qn(t),t)^2}{|\psi(x\qn(t),t)|^2} - J(x\qn(t),t) \dx\qn(t) \right] dt .
\end{align}
The second term in the last integral of \eref{eq.vmax} is orders of magnitude smaller than the first one (and decreases as we increase the adiabaticity) since $\dx\qn$ is much smaller than the peak velocities.

\begin{figure}
\centerline{\includegraphics[width=\columnwidth]{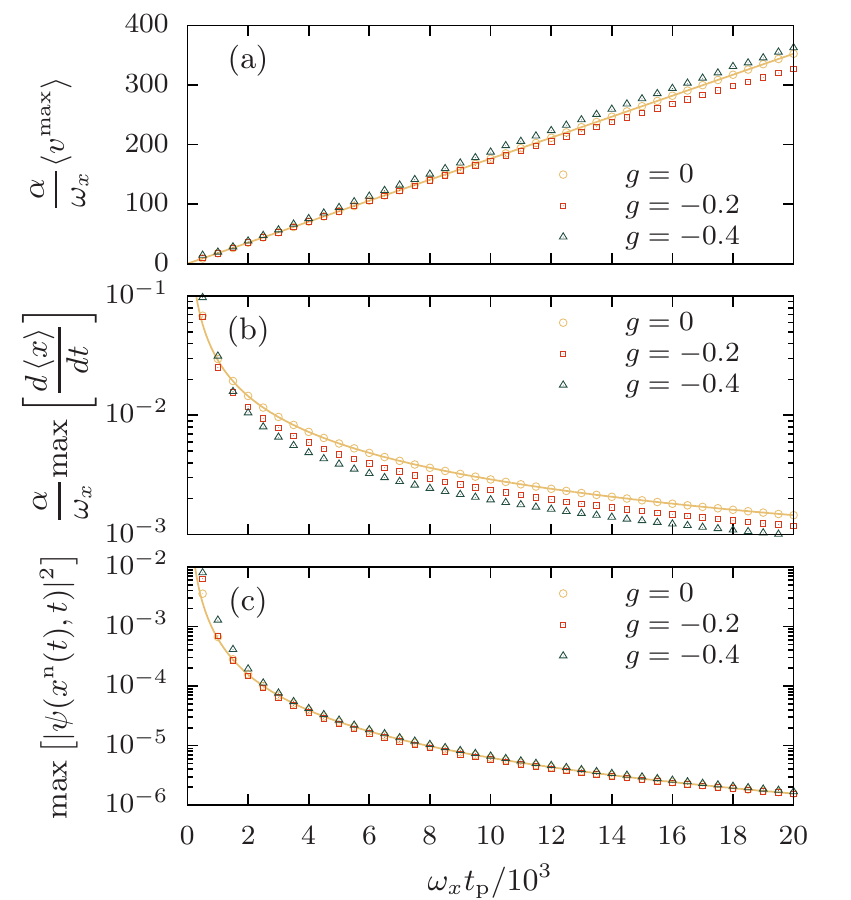}}
\caption{
\label{fig.gamma}
(color online)
Results for different simulations with different values of $t_{\rm p}$ and $g$.
(a) Mean trajectory velocity at the node.
(b) Maximum mean atomic velocity.
(c) Maximum probability density at the node.
Solid lines are, respectively, a linear fit, a $t_{\rm p}^{-1}$ fit and a $t_{\rm p}^{-2}$ fit for the $g=0$ data.
}
\end{figure}

We plot in \fref{fig.gamma} the result of various simulations for different times $t_{\rm p}$ and different interaction strengths $g$.
Variations in $g$ slightly change the values of the calculated quantities, but the overall behavior remains the same.
Thus, for clarity we have performed fits for the $g=0$ data only.
\Fref{fig.gamma}(a) shows that the mean velocity that the trajectories reach when passing through the node grows linearly with $t_{\rm p}$.
Note that all trajectories accelerate abruptly, but at different times, and then, the maximum value that the mean atomic velocity, $\textrm{max}\left[{d\langle x \rangle}/{dt}\right]$, takes during the evolution decreases inversely proportional to $t_{\rm p}$, see \fref{fig.gamma}(b), remaining orders of magnitude below $\langle v^\textrm{max} \rangle$.
There is no apparent limit to the trajectory velocities in the middle region as we approach the limit of perfect adiabaticity,
and at some point they might surpass the speed of light.
Bohmian velocities calculated from the Dirac equation are always bounded by the speed of light~\cite{Superluminal}, implying that relativistic corrections to the Schr\"odinger equation would be needed to properly address the transport while avoiding superluminal matter wave propagation in the adiabatic limit.
It is surprising that the Schr\"odinger equation (or the GPE) cease to be valid and one should consider such corrections in the limit where the process is performed ``infinitely'' slow.
It is also striking that, at variance with the usual tunneling problems where such high-velocities occur at the energetically-forbidden regions, here they appear in a minimum of potential.

\Fref{fig.gamma}(c) shows that the maximum population of the node during the evolution decreases inversely proportional to $t_{\rm p}^2$.
As we already mentioned, this non-zero population appears due to an imperfect following of the dark state, and its decrease is what makes the  high velocities in the middle region to grow, as a classical fluid is forced to accelerate when traveling through a narrowing conduct.
Note that a perfect node would forbid the transport, and thus a non-zero population is needed for the transport to take place.
We have computed the integrated flux through the node, $\int_0^T J(x\qn(t),t)dt$, and as expected, found it to be 1 in all cases, meaning that the entire wavefunction that is transported from the left to the right region is transiting through the (quasi)node in the middle region.

In summary, by using the Bohmian formalism of quantum mechanics applied to the adiabatic transport of a BEC in a triple well we have shown that quantum transport always implies transit.
The absence of atomic population in the middle region is explained by an increase of the trajectories velocities in the vicinity of the node of the followed dark state.
As the process is performed more adiabatically, the trajectories velocities increase with no apparent limit, and thus relativistic corrections should be taken into account to describe correctly the dynamics of the system.
It would be interesting to investigate if the Doppler shift in the light absorption due to the atomic velocities could be used to (weakly) measure the atomic momenta and study some of the features described in this paper in similar lines as in Ref.~\cite{Steinberg}.
Moreover, during the adiabatic transport, the velocity change is very abrupt, leading to very high accelerations (and decelerations).
It remains an open problem to investigate whether the matter wave STIRAP technique for a charged particle could lead to the emission of radiation.

The authors gratefully acknowledge discussions with Alfonso Alarc\'on, Gerhard Birkl, Albert Bram\'on, Maciej Lewenstein and Daniel Viscor,
and financial support through Spanish MICINN contracts  TEC2009-06986, FIS2008-02425, and CSD2006-00019, and the Catalan Government contracts SGR2009-00347 and SGR2009-00783.
Albert Benseny acknowledges financial support through grant AP 2008–01275 from the MICINN.

\end{document}